\documentclass[12pt]{article}
\usepackage{a4}
\usepackage{epsfig}
\usepackage{pslatex}  
\newcommand{\simlt}  {\raisebox{-.6ex}{$\stackrel{\textstyle <}{\sim}$}}

\newcommand{\jcp} {\ensuremath{J}}

\newcommand{\mt}{$\mu$-$\tau$}
\newcommand{\JJ}{J}
\newcommand{\SK}{{\cal S}}

\newcommand{\DD}{{\cal R}}
\newcommand{\DK}{{\cal K}}
\newcommand{\implies} {\Rightarrow}
\newcommand{\rb}{\overline{\rho}}
\newcommand{\nb}{\overline{\eta}}

\newcommand{\beq}{\begin{equation}}
\newcommand{\eeq}{\end{equation}}
\newcommand{\bea}{\begin{eqnarray}}
\newcommand{\eea}{\end{eqnarray}}

\begin{document}
\begin{flushright}
15th January 2009\\
\end{flushright}
\vspace{2 mm}
\begin{center}
{\Large Is the Unitarity Triangle Right?}
\end{center}
\vspace{1mm}
\begin{center}
{P.~F.~Harrison\footnotemark[1] and D.~R.~J.~Roythorne\\
Department of Physics, University of Warwick,\\
Coventry CV4 7AL. UK}
\end{center}
\begin{center}
{and}
\end{center}
\begin{center}
{W.~G.~Scott\\
Rutherford Appleton Laboratory,\\
Chilton, Didcot, Oxon OX11 0QX. UK}
\end{center}
\vspace{1mm}
\begin{abstract}
\baselineskip 0.6cm
The latest fits to the CKM matrix indicate that $\alpha=(90.7^{+4.5}_{-2.9})^{\circ}$. 
The proximity of $\alpha$ to a right-angle raises the question: is it 
merely accidental or is it due to some physics beyond the Standard Model?
In the framework of our recently-proposed flavour permutation symmetry, 
we consider the similarities between the quark and lepton mixing matrices, $V$ and $U$, 
arguing that the relative smallness of one element in each suggests common constraints.
These constraints link the smallness of $V_{ub}$ and 
$U_{e3}$ with each other, and with the approximate 
$\mu-\tau$ symmetry observed in leptonic mixing, together with a prediction of a 
large Dirac $CP$ phase in both the quark and lepton sectors. In the quark case, we predict 
$\alpha=(89.0\pm 0.2)^{\circ}$, in agreement with data and suggesting that the 
unitarity triangle is in fact very nearly, but not exactly right.
\vspace{10mm}
\begin{center}
Talk given at the International Conference on Particles and Nuclei (PANIC08), Eilat, Israel, 13th November 2008.\footnotemark[2]
\end{center}
\end{abstract}

\footnotetext[1]{Speaker and Corresponding Author, E-mail:p.f.harrison@warwick.ac.uk}
\footnotetext[2]{{\tt http://www.weizmann.ac.il/MaKaC/contributionDisplay.py?contribId=235\&session}\\{\tt Id=41\&materialId=slides\&confId=0}}
\newpage
I have to thank Bjorken for the title of my talk, since he posed just this 
question in 1997 \cite{bjHawaii}.
He was interested, together with Stech \cite{stech}, in the possibility that 
the angle $\gamma$ of the unitarity triangle (UT) might be exactly 90$^{\circ}$. He also 
commented (negatively) on an earlier prediction by Fritzsch and Xing \cite{fritzschXing}
that $\alpha\sim 90^{\circ}$ might be preferred. The latest 
fits \cite{CKMUTFIT} to the CKM matrix using data from the $B$ factories and 
other facilities indicate that $\alpha=(90.7^{+4.5}_{-2.9})^{\circ}$, 
so that Fritzsch and Xing may indeed have been correct. Their successful prediction 
was based on their ``four texture zero'' model of hermitian quark mass matrices which 
needed $\alpha$ in the range $\sim 90^{\circ}\pm 20^{\circ}$ to be consistent with the famous 
relationship: $\sin^2\!{\theta_C}\simeq\frac{m_d}{m_s}\!+\!\frac{m_u}{m_c}$. 
We proposed in 2007 \cite{HRS1} an alternative explanation for $\alpha\simeq 90^{\circ}$.

We first introduced the idea \cite{HRS1} of flavour-symmetric mixing observables 
(FSMOs), namely those mixing observables having specific transformation properties 
(either even or odd) under the discrete group of (separate) permutations of the 
rows and columns of the mixing matrix (for either quarks or leptons). FSMOs 
share this flavour-symmetry property with the Jarlskogian \cite{JCP:1}, 
$\jcp$, which is the prototype flavour-odd observable (with $\jcp^2$ a 
prototype flavour-even variable). FSMOs were also shown to be expressible 
in terms of simple functions of the relevant mass matrices.
We now propose \cite{Kbased}, a complete set of such variables based 
on the $K$-matrix \cite{PKQ} of mixing observables, 
$K_{\alpha i}:={\rm Re}\,(V_{\beta j}V_{\gamma k}V^*_{\beta k}V^*_{\gamma j})$,
which is the $CP$-conserving analogue of $\jcp$:
\bea
\SK:=\sum_{\gamma k}K_{\gamma k};&\qquad&
\DD':=\sum_{\gamma k}(K_{\alpha i}K_{\beta j}+K_{\alpha j}K_{\beta i});\cr
\DK:={\rm Det}\,K;&\qquad&
3\JJ^2=\sum_{\alpha}\sum_{i\neq j}K_{\alpha i}K_{\alpha j}
=\sum_{i}\sum_{\alpha\neq \beta}K_{\alpha i}K_{\beta i}.
\label{FSMOs}
\eea
$\SK$, $\DD'$ and $\JJ^2$ are flavour-even, while $\DK$ is odd.
These four variables completely specify the physical content of complex $3\times 3$ 
mixing matrices, modulo permutations of rows and columns and the sign 
of $\JJ$. Expressions for them in terms of the standard (PDG) mixing parameters,
the Wolfenstein parameters and our $P$-matrix based FSMOs \cite{HRS1} are given 
in \cite{Kbased}. We note the following suggestive hierarchy among them: 
i) $CP$-phase $\delta=0\Rightarrow \JJ=0$;
\hbox{ii) Any one mixing angle, $\theta_{ij}=0\Rightarrow \JJ=\DK=0$;} 
iii) Any two $\theta_{ij}=0\Rightarrow \JJ=\DK=\DD'=0$; 
iv) All three $\theta_{ij}=0\Rightarrow \JJ=\DK=\DD'=\SK=0$.

It is often emphasised that the quark and lepton mixing matrices have very 
different forms. However, a unified description remains an important goal, 
which we seek in flavour-symmetric terms \cite{HRS1}. 
Both the quark and lepton mixing matrices share the common feature that their top
right-hand element is significantly smaller than the typical magnitudes of their
other elements. Hence we seek a flavour-symmetric condition for one small mixing 
matrix element. Now, a mixing matrix has a zero element if and only if:
%
\beq
\DK=0 \quad{\rm and}\quad \JJ=0.
\label{constraint2}
\eeq
We recall that with \mt-reflection symmetry \cite{MUTAUSYMM} (which holds at least to 
a good approximation in the leptonic case), the constraint $\DK=0$ follows trivially 
from the elementary properties of determinants, since in that case, the $\mu$ and 
$\tau$ rows of the $K$ matrix are equal to each other. Moreover, both conditions, 
Eq.~(\ref{constraint2}), hold exactly in tri-bimaximal mixing \cite{TBM}.
The two conditions, Eq.~(\ref{constraint2}), may be combined, both being implied 
when the product, ${\cal P}$, of the moduli-squared of all nine elements of the mixing 
matrix satisfies:
\beq
{\cal P}=0,\quad{\rm where}
\quad{\cal P}\equiv\prod_{\alpha i} |V_{\alpha i}|^2=\DK^2+\jcp^2(2\jcp^2+{\cal{R}}')^2,
\label{prodP}
\eeq
which is zero iff $\DK=0$ and $\jcp=0$ (since ${\cal{R}}'\ge 0$ always).

For a small, rather than a zero mixing matrix element, one or other 
(or both) of the conditions, Eq.~(\ref{constraint2}), should be relaxed. Further, 
since for the quarks, clearly $\JJ\neq 0$, while 
for both quarks and leptons, the data are consistent with $\DK=0$, we are led to 
the following predictive conjecture, consistent with both the quark and lepton data:
\beq
\DK=0 \quad{\rm and}\quad |\jcp/\jcp_{max}|={\rm small}
\label{conjecture}
\eeq
(it is not implied that the small quantity necessarily has the same value in both 
sectors). Equation (\ref{conjecture}) is a unified and flavour-symmetric, partial 
description (in the sense that only two degrees of freedom are constrained)
of both lepton and quark mixing matrices. It implies the existence of at least one small element in each mixing matrix, as manifested by $U_{e3}$ and $V_{ub}$, and is associated with $\mu-\tau$ symmetry \cite{MUTAUSYMM}, which is suggested by the lepton data.

While relaxing slightly {\em both} constraints, Eq.~(\ref{constraint2}), allows a small 
mixing angle, \hbox{$\theta$ (eg.~$\theta_{13}$),} and an arbitrary $CP$-phase, relaxing 
instead {\em only} the $\jcp$-constraint while keeping \hbox{$\DK=0$}, Eq.~(\ref{conjecture}), 
still yields a small mixing angle, $\theta$, but now subject to the 
phase-condition that one UT angle $\rightarrow 90^{\circ}$, as $\jcp\rightarrow 0$
[such that $(\theta/\theta_{ij})\rightarrow 0$ (for $\theta_{ij}\neq\theta$)]. Moreover, 
since the $(\theta_{13}/\theta_{ij})$ {\em are} small for both 
the quarks and leptons, the deviations from $90^{\circ}$ are small and calculable in terms 
of them. Flavour symmetry prevents an a priori prediction of {\it which} UT angle is 
$\simeq 90^{\circ}$, but this can be obtained from the data (in general, flavour symmetry 
is spontaneously broken by the solutions of flavour-symmetric constraint equations).

For the quarks, we use our formula \cite{Kbased, ANGLES} for $\DK$ in the Wolfenstein 
parameterisation (being based on data, the latter already breaks flavour symmetry) 
to show that \hbox{$\DK=0\implies \alpha \simeq 90^{\circ}$}. Defining the squares of 
the two non-trivial sides of the standard $B$ physics unitarity triangle as 
\hbox{$u\equiv\overline{\rho}^2+\overline{\eta}^2$} and 
$v\equiv 1-2\overline{\rho}+\overline{\rho}^2+\overline{\eta}^2$, 
the cosine rule gives $\cos{\alpha}=\frac{u+v-1}{2\sqrt{uv}}$. 
Meanwhile $\DK=0$ implies \cite{Kbased, ANGLES}:
\beq
u+v-1
\equiv 2(\rb^2+\nb^2-\rb)
=2\lambda^2\,u\,v\Rightarrow \cos{\alpha}=\lambda^2\sqrt{u\,v}
\simeq\lambda^2\overline{\eta}
\label{cosAlpha}
\eeq
at leading order in $\lambda^2$ (it vanishes as 
$\frac{\sin\theta_{13}}{\sin\theta_{23}}\sim\lambda\sqrt{u}\rightarrow 0$, 
as foreseen). Eq.~(\ref{cosAlpha}) predicts:
\beq
(90^{\circ}-\alpha)\simeq\lambda^2\overline\eta=1.0^{\circ}\pm 0.2^{\circ},
\label{qPrediction}
\eeq
compared with its current measured value \cite{CKMUTFIT}:
%
$(90^{\circ}-\alpha)=2^{\circ\,+5^{\circ}}_{\,\,\,\,-6^{\circ}}$
%
(direct measurement) or $(90^{\circ}-\alpha)=-1^{\circ\,+3^{\circ}}_{\,\,\,\,-5^{\circ}}$ 
(full CKM fit). The unitarity triangle is indeed, very nearly right!
It will be interesting to test Eq.~(\ref{qPrediction}) more precisely in future 
$B$ physics experiments, such as LHCb and super flavour factories.

We know experimentally that for leptons, $U_{e3}$ is the only small mixing matrix element.
Hence, only the UT angles $\phi_{\mu 1}$, $\phi_{\tau 1}$, $\phi_{\mu 2}$ and 
$\phi_{\tau 2}$ (using the nomenclature of \cite{ANGLES, BHS05}) 
can be close to $90^{\circ}$, and they all satisfy (see eg.~Fig.~1 of \cite{BHS05})
$|\cos{\phi_{\alpha i}}|\simeq |\cos{\delta}|$, where \hbox{$\delta={\rm Arg}(U_{e3})$}. 
Using our formula \cite{Kbased} for $\DK$ in terms of PDG parameters, 
we find that $\DK=0$ implies:
\beq
|90^{\circ}-\delta|\simeq 4\cot{2\theta_{12}}\cot {2\theta_{23}}\sin{\theta_{13}}
\simeq 2\sqrt{2}\,\sin{\theta_{13}}\,\sin{(\theta_{23}-45^{\circ})}\,\simlt\, 4^{\circ},
\eeq
($1\sigma$) at leading order in small quantities (again vanishing as 
$\frac{\sin\theta_{13}}{\sin\theta_{23}}\rightarrow 0$ as expected).
Our conjecture thus predicts a large 
$CP$-violating phase in the MNS matrix, which is promising for the discovery of leptonic $CP$ 
violation at eg.~a future Neutrino Factory.
%
%
%
%

%
\end{document}